\documentclass[sigconf]{acmart}

\usepackage[most]{tcolorbox}
\usepackage{graphicx}
\usepackage{listings}
\usepackage{xcolor}

\AtBeginDocument{%
  }

\setcopyright{acmcopyright}
\copyrightyear{2024}
\acmYear{2024}
\acmDOI{XXXXXXX.XXXXXXX}

\acmConference[ICSE 2024]{46th International Conference on Software Engineering}{April 2024}{Lisbon, Portugal}

\begin{document}

\title{Towards Automatic Translation of Machine Learning Visual
Insights to Analytical Assertions}


\author{Arumoy Shome}
\email{a.shome@tudelft.nl}
\affiliation{%
  \institution{Delft University of Technology}
  \country{Netherlands}
}

\author{Lu{\`\i}s Cruz}
\email{l.cruz@tudelft.nl}
\affiliation{%
  \institution{Delft University of Technology}
  \country{Netherlands}}

\author{Arie van Deursen}
\email{arie.vandeursen@tudelft.nl}
\affiliation{%
  \institution{Delft University of Technology}
  \country{Netherlands}
}

\begin{abstract}

We present our vision for developing an automated tool capable of
translating visual properties observed in Machine Learning (ML)
visualisations into Python assertions. The tool aims to streamline the
process of manually verifying these visualisations in the ML
development cycle, which is critical as real-world data and
assumptions often change post-deployment. In a prior study, we mined
$54,070$ Jupyter notebooks from Github and created a catalogue of
$269$ semantically related visualisation-assertion (VA) pairs.
Building on this catalogue, we propose to build a taxonomy that
organises the VA pairs based on ML verification tasks. The input
feature space comprises of a rich source of information mined from the
Jupyter notebooks---visualisations, Python source code, and associated
markdown text. The effectiveness of various AI models, including
traditional NLP4Code models and modern Large Language Models, will be
compared using established machine translation metrics and evaluated
through a qualitative study with human participants. The paper also
plans to address the challenge of extending the existing VA pair
dataset with additional pairs from Kaggle and to compare the tool's
effectiveness with commercial generative AI models like ChatGPT. This
research not only contributes to the field of ML system validation but
also explores novel ways to leverage AI for automating and enhancing
software engineering practices in ML.

\end{abstract}


\maketitle

\section{Introduction}

Visualisations are employed at various stages of a Machine Learning
(ML) pipeline. They are used to understand and verify data properties
during the early stages, summarise metrics and fine-tune models during
development, and monitor performance post-deployment. The iterative
and experimental nature of building ML systems heavily relies on
insights from visualisations to guide design and implementation
decisions~\cite{yuan2020survey,hohman2019visual,amershi2015modeltracker,wexler2019what-if,shome2022data,haakman2021ai}.

However, real-world data that ML systems encounter post- deployment
seldom remain static. They often change as a reflection of the world,
potentially violating initial assumptions made during
development~\cite{amershi2019software,lwakatare2021on}. Every
subsequent iteration of the ML development cycle used to retrain and
update the ML model, therefore demands manual validation of the
visualisations that were used to test ML system properties.

Assertions or analytical tests derived from ML visualisations can
significantly reduce manual verification efforts. Such formal
assertions record the AI practitioner's observations about the model
or data at a specific moment. They also serve as a reference point for
future AI practitioners to understand the interpretations made from
earlier visualisations.

In a prior study currently under review, we investigate how
frequently analytical tests are formulated from visualisations in ML
systems and analysed the effectiveness of these tests. We mined
$54,070$ Jupyter notebooks from GitHub that contain assertions written
in Python and developed a novel methodology to identify $1,764$
notebooks which contain an assertion below a visualisation. We
manually analysed the $1,764$ notebooks and catalogued $269$
semantically related visualisation-assertion (VA) pairs (to be
released publicly with the camera-ready version of the paper). Further
in-depth analysis of the VA pairs revealed that assertions often fail
to capture all the information obtained from the corresponding
visualisations. Our results indicate that formulating analytical
assertions from visualisations is an emerging testing technique in ML.
However, it also highlights the limitations of current practices,
demonstrating a need for automated tools that can assist ML
practitioners in validating visualisations and formulating more
comprehensive analytical assertions.

In this paper, we propose our research vision to develop an automated
tool to generate analytical assertions from ML visualisations. The
research questions we plan to address along with the contributions we
envision to make are presented below.

\begin{description}
  \item[RQ1.] \textbf{How are VA pairs used to perform ML verification
    tasks?}

    We begin by creating a taxonomy that clusters the existing $269$
    VA pairs by specific ML verification tasks. This taxonomy not only
    enables us to fine-tune the data and model for the automated tool,
    but also serves as valuable reference for future ML practitioners.

  \item[RQ2.] \textbf{What kind of input features enable AI models to
    generate assertions from visualisations?}

    The dataset consists of visualisations, source code and natural
    text. Based on the manual analysis we previously conducted to
    identify semantically related VA pairs, we hypothesis that the
    source code for the visualisation is sufficient for ML models to
    generate meaningful assertions. We plan to compare the results to
    multi-modal input feature space where a combination of image, code
    and text are used.

  \item[RQ3.] \textbf{What kind of AI models generate the best
    assertions from visualisations?}

    A empirical study will be conducted to compare traditional
    NLP4Code models with modern Large Language Models. The
    effectiveness of these models will be evaluated both using
    a holdout set and additionally using a qualitative study with
    human participants.

  \item[RQ3.] \textbf{How does our tool compare to commercial
    Generative AI models?}

    This paper presents results from a preliminary study which we
    conducted to evaluate the effectiveness of the ChatGPT 4.0 model
    to automatically generate assertions from ML visualisations. We
    plan to extend the preliminary study to the full catalogue on VA
    pairs and compare the results to our proposed tool.

\end{description}

\section{Our Vision}

\begin{figure*} \includegraphics[width=\textwidth]{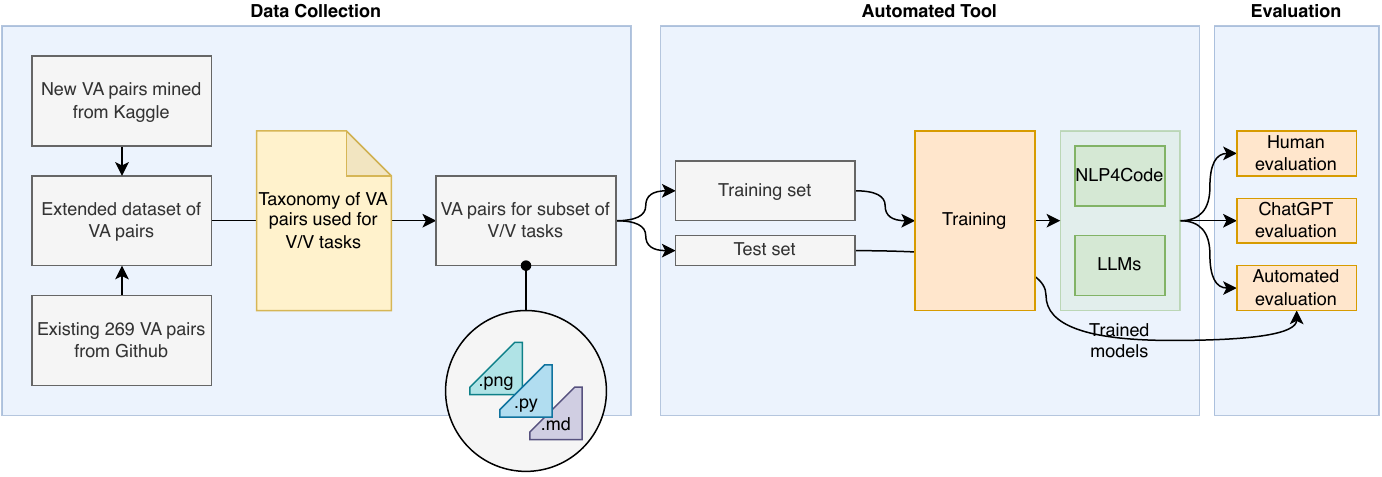}
\caption{Vision for the automated tool proposed in this paper to
generate analytical assertions from ML visualisations.}
\label{fig:vision} \end{figure*}

Figure~\ref{fig:vision} provides a visual summary of the research
vision proposed in this paper. During the data collection phase, an
extended catalogue of VA pairs will be created using VA pairs mined
from Github and Kaggle. A taxonomy of VA pairs used for ML
verification tasks will be created and used to fine-tune the dataset
and model. Quantitative studies will determine the best input feature
space and AI models that are able to generate meaningful assertions
from visualisations. Quantitative and qualitative evaluations will be
employed to position the proposed tool with-respect-to ChatGPT.

\subsection{RQ1: How are VA pairs used to perform ML verification tasks?}

The existing catalogue of $269$ VA pairs was created by mining public
Jupyter notebooks on Github, that contained both visualisations and
assertions. We searched the entirely of Github and obtained $54,070$
Jupyter Notebooks written in Python, containing the keyword ``assert''
in them. The sample was further refined to $1,764$ notebooks that
imported popular Machine Learning or Data Science libraries and had
specific cell arrangements that indicated a potential relationship
between visualisations and assertions. We manually reviewed these
notebooks to identify $269$ VA pairs that were semantically related to
each other. This analysis involved manually examining the source code
of the visualisation and assertion cells and excluding unrelated
pairs.

The $269$ VA pairs however lack structure. Therefore, We plan to
organise them using a taxonomy that categories VA pairs according to
specific ML verification tasks. This taxonomy is critical for the
creation of an automated tool, as it allows for more targeted
fine-tuning of the training data and model. We hope that by aligning
the model with distinct verification tasks identified in the taxonomy,
the tool can generate more accurate and relevant assertions.

The taxonomy also organises the domain knowledge of ML visualisations
and assertions. It can be used by future ML practitioners as reference
to identify the types of visualisation that can be used for a given
verification task, or formulate comprehensive assertions given a visualisation.

\subsection{RQ2: What kind of input features enables AI models to
generate assertions from visualisations?}

Using the taxonomy created in the prior phase, we plan to create
a high quality dataset of VA pairs which can be used to train and test
AI models that automatically generate the assertions. The input
feature space comprises of three rich sources of information---the
visualisations stored as PNG images, Python source code used to create
the visualisation and when available, natural text written by the
notebook authors above and below the visualisation.

In our prior study, we analysed $1,764$ notebooks manually to identify
$269$ VA pairs. We reflected on our manual analysis procedure and
realised that the code for the visualisation and the assertion provide
sufficient context to determine whether they are related to each other
or not. Our hypothesis is that the Python source code along with the
Markdown text should provide sufficient information for the ML model
to generate meaningful assertions. However, it will also be
interesting to compare the results with multi-modal AI models which
also use the visualisations as input.

\subsection{RQ3: What kind of AI models generate the best assertions
from visualisations?}

We plan to conduct an extensive empirical study to determine what type
of AI models are able to produce the best assertions. Here, we plan to
compare and contrast traditional NLP4Code models with open-source
Large Language Models (LLMs).

From an NLP perspective, creating assertions given the source code for
a visualisation is a translation task. Thus, the effectiveness of the
above models can be quantified using existing translation metrics such
as \textit{BLEU}, \textit{Meteor} and
\textit{Rouge-L}~\cite{zhu2023revisiting}.

We further plan to evaluate the practical relevance and impact of our
proposed tool by conducting a qualitative study with human
participants. We are currently exploring two approaches here. First,
with human participants that compare the ground truth assertions with
the ones generated by our tool, and rate them on a Likert Scale on
Naturalness and Correctness~\cite{izadi2022on}. An alternative
approach would be to collect feedback from the notebook authors by
opening Pull Requests on Github that propose the generated assertion.

\subsection{RQ4: How does our solution compare to commercial
generative AI models?}

The ``Elephant in the room'' must be addressed. OpenAI's suite of
commercial generative AI models has produced a tectonic shift in the
Software Engineering landscape. We plan to evaluate the effectiveness
of our proposed tool against commercial AI models such as ChatGPT and
DALL-E.

In a preliminary study, we assessed the capability of ChatGPT 4.0 to
generate assertions from visualisations, utilising six VA pairs that
were highlighted in our prior study. The study employed one-shot
prompting with three different inputs---providing ChatGPT with only
the visualisation image, only the code generating the visualisation,
and both the image and code. The responses were evaluated based on
three criteria---whether the generated assertions reflected properties
or trends evident in the visualisations, the similarity of the
response to the ground truth assertions, and whether ChatGPT proposed
alternative assertions that validated the same visual properties as
the ground truth.

All 18 responses from ChatGPT contained assertions derived from visual
properties, notably even when prompted solely with the code. Out of
these, 8 responses closely matched the ground truth assertions, and in
the remaining 10, ChatGPT suggested valid alternative assertions
aligned with the same visual properties as the ground truth. We plan
to extend this study to the entire catalogue of VA pairs which will
allow us to compare and contrast our proposed tool to the abilities of
ChatGPT.

\section{Challenges}

\subsection{Dataset Limitations and Bias}

The quality and diversity of the dataset used to train the tool will
heavily influence its effectiveness. There is a risk of bias if the
datasets are not representative of real-world scenarios or if they
lack diversity in terms of the types of visualisations and assertions.

In RQ1, the existing $269$ VA pairs were obtained from Github which
might not be representative of all VA pairs observed in ML. To
mitigate this threat, we plan to extend the catalogue with VA
pairs mined from Kaggle. This also allows us to simplify our data
mining process by eliminating the need to filter notebooks for
relevant libraries since Kaggle is primarily targeted towards
the Data Science and Machine Learning communities.

In RQ2, we expect to face challenges with respect to the quantity and
quality of the training data. Since translating visual properties to
analytical assertions is an emerging testing practise, we may not have
sufficient training data for our AI models. The extension using VA
pairs from Kaggle should alleviate this threat to a certain extent.
Another alternative would be to fine-tune the NLP4Code models using
our catalogue. With LLMs, we can experiment with different prompting
techniques such as few-shot prompting~\cite{liu2023pretrain}.

With respect to quality, we want to reinforce the manual analysis
required to determine whether a given visualisation and assertion are
related, with multiple annotators. The quality of the annotations can
be determined by calculating the inter-rater agreement using the
\textit{Kappa} metric.

In RQ4, we do not know the contents of the training corpus used to
train ChatGPT. However given its colossal size, it is safe to assume
that our catalogue of VA pairs exists in its training data (or will be
included in future versions). This supports our decision to use
open-source LLMs where we can validate that no such data leakage
occurs. We can also extend the proposed evaluation for our tool (RQ3)
and ChatGPT (RQ4) using VA pairs that have not been publicly released.
One potential solution here would be to collect VA pairs created by
students in a Software Engineering for ML course.

\subsection{Complexity of Visual Analytics}

Translating visual properties accurately into analytical assertions is
a complex task. The tool must be capable of understanding and
interpreting a wide range of visual patterns and nuances, which is
a significant challenge. In our prior paper, we identified several
instances where the analytical assertions were not able to capture all
the information present in their visual counter-part. While our
proposed tool may not be able to capture all the nuances of
a visualisation every time, identifying this boundary of automation
and human intervention remains an important and exciting research
direction.

\subsection{Integration into Existing Workflows}

For the tool to be widely adopted, it needs to be seamlessly
integrated into existing ML development workflows. This includes
compatibility with various programming environments and ML frameworks.
While the outcome of our research endeavours may be a simple
prototype, we hope to lay the foundations that enable future
researchers to integrate our tool into existing computational notebook
environments such as Jupyter Lab or Visual Studio Code. Such an
integration may enable the tool to provide better assertions using
additional context either obtained directly from the user through
a chat interface~\footnote{Such as the one provided by Github Co-Pilot
extension for Visual Studio Code:
https://code.visualstudio.com/blogs/2023/03/30/vscode-copilot} or
derived from the entire software codebase~\footnote{One example of
such an editor extension is Cody which is available for Visual Studio
Code and JetBrains: https://github.com/sourcegraph/cody}.

\section{Expected Outcomes}

The proposed tool automates the translation of visual properties from ML
visualisations into Python assertions. This automation reduces the
manual effort required in the model validation process, making it
faster and more efficient. By automating a critical part of the ML
development process, the tool can streamline the entire cycle from
data preprocessing to model deployment, enabling quicker iterations
and enhancements. Automated assertions make it easier to handle large
datasets and complex models, as it removes the bottleneck of manual
analysis, thereby enhancing scalability.

Manual processes are prone to human errors, such as overlooking subtle
visual cues in data. In contrast, an automated tool can consistently
capture and translate these nuances, reducing the likelihood of
errors. This form of automation leads to a more standardized approach
to validating ML models, ensuring that the same criteria are applied
consistently across different models and datasets. Furthermore,
automated assertions provide a reliable record of the state of the
data and model at various points in time, aiding in historical
analysis and model auditing.

An automated tool also enables ML practitioners to adapt assertions as
real-world data evolves. By providing immediate feedback on visual
data, the tool will enable faster updates and refinements to ML models
in response to changing data trends. The tool could potentially
identify anomalies or shifts in data patterns sooner than manual
processes. This not only allows for proactive adjustments to models,
but also significantly reduces effort required to maintain the
relevance and accuracy of ML models over time.

\section{Expected Outcomes Beyond ML Testing}

As ML augments software systems in safety-critical domains, there has
been a significant emphasis on the explainability of ML models. The
complexity and multi-dimensionality of such models and their
underlying data has lead to the adoption of Visual Analytics by the ML
community. However, interpretation of a visualisation may vary between
developers with varying domain expertise and background. Converting
such visual interpretations into assertions can help bring consistency
into the interpretability of ML visualisations across all project
members.

The concept of translating visual properties to formal assertions can
be extended to other domains (such as scientific computing and
simulation), where visual analytics also plays a prominent role in the
decision making process. Future research could explore ways to
customise these tools for specific industries or types of data,
increasing their applicability and effectiveness.

The effectiveness of the tool in real-world scenarios could also open
research into how human practitioners and AI tools can best
collaborate, particularly in the context of complex problem-solving
tasks like ML model validation.

\section{Conclusion}

Developing an automated tool to convert visual properties from ML
visualisations into analytical assertions marks a timely advancement
in the domains of Machine Learning and Software Engineering. This tool
has the potential to improve the efficiency, accuracy, and consistency
of ML model testing and validation. Nevertheless, it also presents
considerable challenges, especially in areas such as visual analysis,
data set quality, and integration with current workflows. The research
initiates various new opportunities, such as enhancing the
interpretability of AI, enabling automated testing, and exploring the
dynamics of human-AI collaboration.

\bibliographystyle{ACM-Reference-Format} \bibliography{bibliography}

\end{document}